\documentclass[preprint,12pt,authoryear,appendixfloats]{aastex61}

\usepackage{float}
\usepackage{lineno}

\begin{document}

\title{Thermal Emission from the Uranian Ring System}

\author{Edward M. Molter}
\affiliation{Astronomy Department, University of California, Berkeley; Berkeley CA, 94720, USA}
\correspondingauthor{Edward Molter}
\email{emolter@berkeley.edu}

\author{Imke de Pater}
\affiliation{Astronomy Department, University of California, Berkeley; Berkeley CA, 94720, USA}

\author{Michael T. Roman}
\affiliation{Department of Physics \& Astronomy, University of Leicester, University Road, Leicester, LE1 7RH, UK}

\author{Leigh N. Fletcher}
\affiliation{Department of Physics \& Astronomy, University of Leicester, University Road, Leicester, LE1 7RH, UK}

%

\begin{abstract}
The narrow main rings of Uranus are composed of almost exclusively centimeter- to meter-sized particles, with a very small or nonexistent dust component; however, the filling factor, composition, thickness, mass, and detailed particle size distribution of these rings remain poorly constrained. Using millimeter (1.3 - 3.1 mm) imaging from the Atacama Large (sub-)Millimeter Array and mid-infrared (18.7 $\mu$m) imaging from the Very Large Telescope VISIR instrument, we observed the thermal component of the Uranian ring system for the first time.
The $\epsilon$ ring is detected strongly and can be seen by eye in the images; the other main rings are visible in a radial (azimuthally-averaged) profile at millimeter wavelengths. A simple thermal model similar to the NEATM model of near-Earth asteroids is applied to the $\epsilon$ ring to determine a ring particle temperature of $77.3 \pm 1.8$ K. The observed temperature is higher than expected for fast-rotating ring particles viewed at our observing geometry, meaning that the data favor a model in which the thermal inertia of the ring particles is low and/or their rotation rate is slow. The $\epsilon$ ring displays a factor of 2-3 brightness difference between periapsis and apoapsis, with $49.1 \pm 2.2$\% of sightlines through the ring striking a particle. These observations are consistent with optical and near-infrared reflected light observations, confirming the hypothesis that micron-sized dust is not present in the ring system.

\end{abstract}

\section{Introduction}

To date, observational data for the Uranian ring system were obtained from a combination of Earth-based stellar occultation measurements \citep[][and references within]{french91}, visible-light and radio occultation data from Voyager 2 \citep[e.g.,][]{smith86, tyler86, gresh89}, visible-light HST \citep[e.g.,][]{karkoschka01a}, and ground-based near-IR observations with adaptive optics \citep[e.g.,][]{depater06, depater07, depater13, dekleer13}. Together, these observations revealed a complex system of ten narrow rings, three broad dusty rings, and at least thirteen associated small satellites \citep[][]{smith86, karkoschka01c, showalter06}. The broad spectral coverage of these observations permitted inferences about the particle size distribution of the rings, showing that the narrow rings (except the $\lambda$ ring) are composed primarily of centimeter- to meter-sized particles, with a very small or nonexistent dust component \citep[][]{gresh89, karkoschka01b}. The $\epsilon$ ring, the brightest and most massive of the narrow rings, was shown to maintain an appreciable eccentricity ($e = 0.00794$) and an azimuthally-varying width; the ring is five times wider and $\sim$2.5 times brighter in reflected sunlight at apoapsis than at periapsis \citep[][]{french88, karkoschka01a}. However, many fundamental parameters about the ring system remain unknown, including the filling factor, composition, thickness, mass, and detailed particle size distribution of each ring \citep[see review in][]{nicholson18}.

In this Letter we present the first millimeter and mid-infrared observations of the Uranian ring system, obtained in 2017 and 2018 with ALMA at three wavelengths (Band 3; 3.1 mm, Band 4: 2.1 mm, Band 6: 1.3 mm) and the VISIR instrument \citep{lagage04} Q2 filter (18.72 $\mu$m) on the Very Large Telescope (VLT). In contrast to images at visible/near-infrared wavelengths, which show the rings in reflected sunlight, our ALMA and VISIR images detect thermal emission from the rings, a component that has never before been imaged. Any contribution from scattered light from Uranus or the Sun is $<$0.01\% (see Appendix \ref{section_rt}). Our observations and data reduction procedures are presented in Section \ref{section_data}. We describe how total flux measurements were derived from radial profiles of the rings in Section \ref{section_radialint}, and then compare these total flux measurements to a thermal model of the $\epsilon$ ring in Section \ref{section_model}. Finally, we present the measured azimuthal structure of the $\epsilon$ ring's thermal component in \ref{section_az} before summarizing our results in Section \ref{section_summary}.

\section{Observations and Data Reduction}
\label{section_data}

\begin{figure}
	\includegraphics[width = 1.0\textwidth]{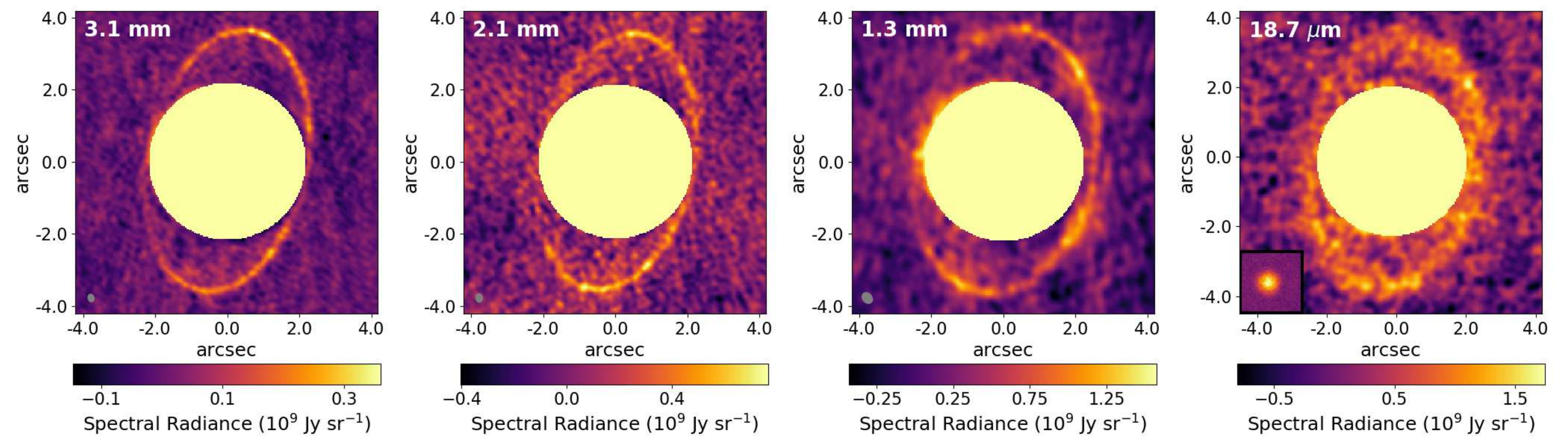}
	\caption{Images of the Uranian ring system at 3.1 mm (ALMA Band 3; 97.5 GHz), 2.1 mm (ALMA Band 4; 144 GHz), 1.3 mm (ALMA Band 6; 233 GHz), and 18.7 $\mu$m (VLT VISIR; 100 THz). The synthesized beams of the ALMA images are shown as grey ellipses in the bottom left corner of each image, and an image of a point source is shown in the bottom left corner of the VLT 18.7 $\mu$m image. The planet itself is masked since it is very bright compared to the rings.\label{prettypics}}
\end{figure}

We observed the Uranian ring system at millimeter wavelengths with the Atacama Large (sub-)Millimeter Array (ALMA) and at mid-infrared wavelengths with the VISIR instrument on the VLT between December 2017 and September 2018.  An observing log is shown in Table \ref{table_obs}, and descriptions of our data processing procedures are given in Sections \ref{subsection_alma} and \ref{subsection_vlt} for the ALMA and VLT data, respectively. Images of the ring system at all four observed wavelengths are shown in Figure \ref{prettypics}; each clearly detects the massive $\epsilon$ ring.

\begin{table}
	\footnotesize
	\begin{tabular}{|c|c|c|c|c|c|c|}
	 & UT Date \& & On-Source & Minimum & Maximum &  Flux \& Gain & Phase \\
	Wavelength & Start Time & Time (min) & Baseline (m) & Baseline (m) & Calibrator & Calibrator \\ 
	\hline
	3.1 mm & 2017-12-03 23:52 & 42 & 41.4 & 5200 & J0238+1636 & J0121+1149 \\
	3.1 mm & 2017-12-06 23:42 & 42 & 41.4 & 3600 & J0238+1636 & J0121+1149 \\
	2.1 mm & 2017-12-27 22:56 & 22 & 15.1 & 2500 & J0238+1636 & J0121+1149 \\
	1.3 mm & 2018-09-13 07:02 & 24 & 15.1 & 1200 & J0237+2848 & J0211+1051 \\
	18.72 $\mu$m & 2018-09-04 08:14 & 42.3 & --- & --- & HD009692 \& HD013596 & --- \\
	18.72 $\mu$m & 2018-10-03 05:18 & 42.3 & --- & --- & HD008498 \& HD010380 & --- \\
	18.72 $\mu$m & 2018-10-13 05:28 & 42.3 & --- & --- & HD011353 \& HD040808 & --- \\
	\end{tabular}
	\caption{Table of observations.\label{table_obs}}
\end{table}

\subsection{ALMA Data}
\label{subsection_alma}

\begin{figure}
	\includegraphics[width = 0.6\textwidth]{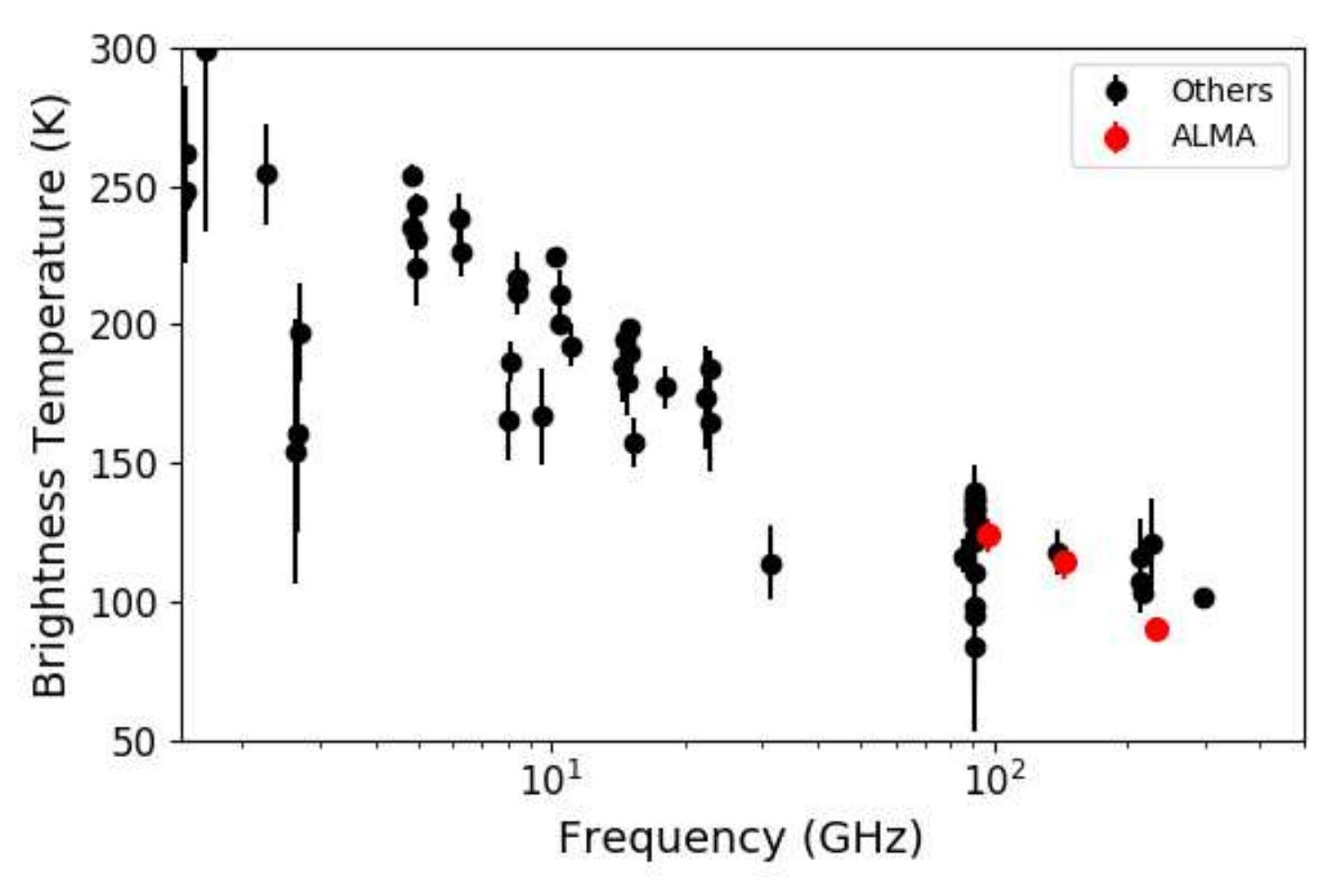}
	\caption{ALMA-derived (red; this paper) and literature \citep[black;][]{gulkis84, depater88} measurements of the average millimeter- and radio-wavelength brightness temperature of Uranus's disk.\label{planetFlux}}
\end{figure}

The data in each of the three bands were flagged and calibrated by the North American ALMA Science Center using the standard data reduction procedures contained in the NRAO's CASA software version 5.1.1. Standard flux and phase calibration procedures were carried out by applying the pipeline using the quasars listed in Table \ref{table_obs} as calibrator sources. The CASA pipeline retrieved flux calibration errors of 5.0\%, 5.4\%, and 5.3\% at 3.1 mm, 2.1 mm, and 1.3 mm, respectively. The flux calibration was sanity-checked by comparing the measured total flux of Uranus to previous millimeter-wavelength observations; the result of this exercise is shown in Figure \ref{planetFlux}. To improve the quality of the map, iterative phase-only self-calibration was performed using a procedure similar to that outlined in \citet{brogan18} using solution intervals of 20, 10, 5, and 1 minutes in that order. To reduce ringing in the image plane from the presence of a bright planet with sharp edges, a uniform limb-darkened disk model of Uranus was subtracted from the data in the UV plane, as done in e.g. \citet{depater14, depater16}. The disk-subtracted data were inverted into the image plane and deconvolved using CASA's \texttt{tclean} function. The ring geometry at the time of observation in each band, taken from JPL Horizons\footnote{https://ssd.jpl.nasa.gov/horizons.cgi} and the Planetary Ring Node,\footnote{https://pds-rings.seti.org/} are given in Table \ref{table_geom}.

\begin{table}
	\begin{tabular}{|c|c|c|c|c|c|}
	Wavelength & d (AU) & $B_{obs}$ (degrees) & $\Omega$ (degrees) & $w$ (degrees) \\ 
	\hline
	3.1 mm & 19.2 & 36.8 & 192 & 337 - 341 \\
	2.1 mm & 19.6 & 36.4 & 192 & 10 \\
	1.3 mm & 19.1 & 43.7 & 190 & 3 \\
	18.72 $\mu$m & 18.9-19.2 & 42.7 - 43.9 & 190 & 351 - 44 \\
	\end{tabular}
	\caption{Observing geometry of the Uranian ring system at the time of our observations. $\Omega$ denotes the longitude of the ring plane ascending node, and $w$ denotes the argument of periapsis of the $\epsilon$ ring. Ranges of values are shown for the 3.1 mm and 18.7 $\mu$m observations because they were taken over the course of several nights.\label{table_geom}}
\end{table}

\subsection{VLT Data}
\label{subsection_vlt}

The 18.72 $\mu$m (Q2 filter) VISIR image combined observations from three nights in September and October 2018; details of these observations are listed in Table \ref{table_obs}. Data from each night were reduced with standard infrared chopping and nodding techniques using the European Southern Observatory VISIR pipeline, and resulting images were flux calibrated via comparison to observed standard stars using custom-written \texttt{IDL} routines. The three calibrated images were then combined and weighted by the inverse of their errors squared to yield the final, absolutely calibrated mean image. Random errors were estimated from the standard deviation of the background sky, and we estimate a 20\% systematic error in radiance due to the uncertainty in the stellar flux and sky subtraction \citep[][]{dobrzycka08}.

\section{Radial Profiles and Total Flux Measurements}
\label{section_radialint}

\begin{figure}
	\includegraphics[width = 0.5\textwidth]{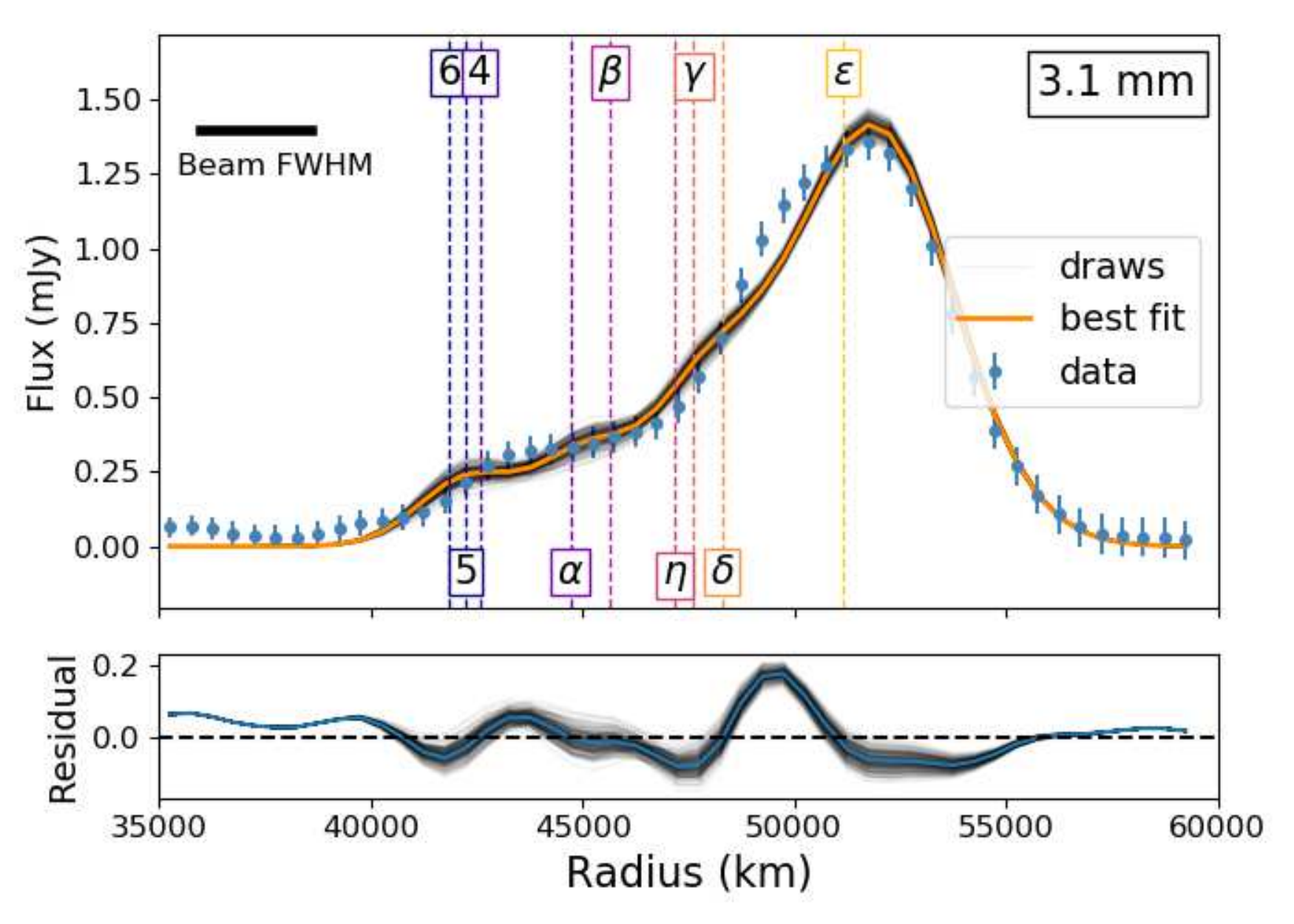} \\
	\includegraphics[width = 0.5\textwidth]{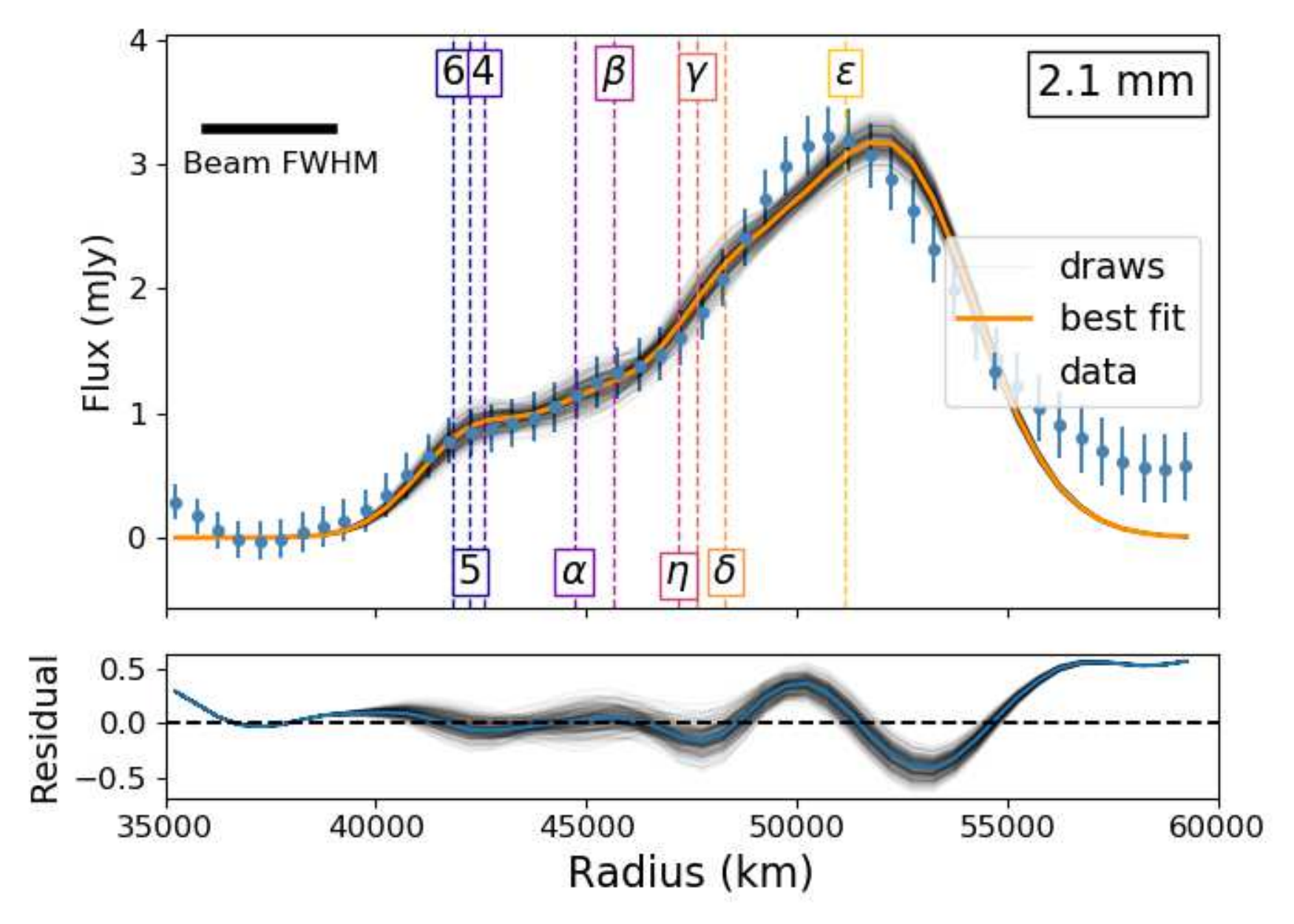} \\
	\includegraphics[width = 0.5\textwidth]{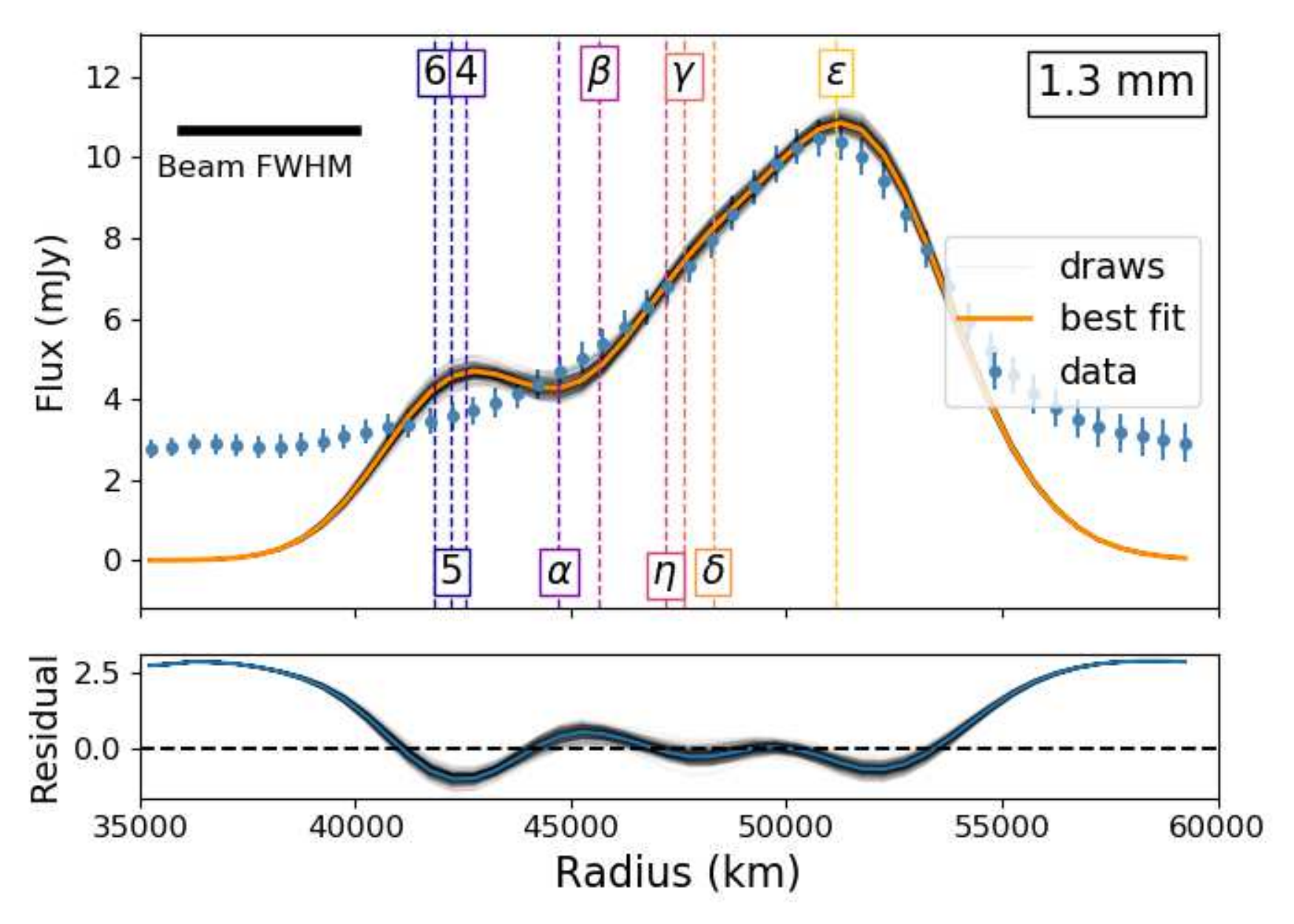}
	\caption{Radial flux profiles of the Uranian ring system at 3.1 mm, 2.1 mm, and 1.3 mm. The data, binned into 500 km intervals, are shown as blue dots. The best-fitting model is shown as an orange line, and draws from the distribution of allowed models are shown as thin grey lines. Locations of the main rings are labeled. \label{radialint}}
\end{figure}

We measured the radial profile of the rings in each ALMA image by integrating the rings in azimuth as follows. First, the images were divided into elliptical annuli using the \texttt{astropy}-affiliated \texttt{photutils} Python package \citep[][]{bradley19}. Each annulus was centered at the center of Uranus and given an eccentricity and angle of rotation such that it mimicked a circular annulus projected into Uranus's ring plane. Each annulus was given a ring-plane width of 500 km, and the ring-plane radius of the annulus inner edge was varied in steps of 500 km from 35000 km to 64500 km. The planet Uranus was masked out to a distance equal to its mean radius plus three times the ALMA full beam width at half power in each observing band. The \texttt{photutils} package was used to compute the geometric fraction of each pixel contained within a given annulus, weight the flux within that pixel according to that fraction, then sum up the weighted flux of all unmasked pixels.  This provided a total flux measurement within the exact elliptical region specified by each annulus (ignoring any masked pixels). The result was a radial flux profile of the ring system in which each unmasked pixel was counted exactly once total, splitting its flux between two annuli if it fell on an annulus edge. The error $\sigma_n$ on each flux measurement was taken to be the RMS noise in the image (computed far from the planet and rings) times the area of the annulus (not including any masked pixels) in units of the synthesized beam area. The observed profile is shown in Figure \ref{radialint}.

The radial flux profile was modeled as the sum of flux contributions from the 6/5/4, $\alpha$/$\beta$, $\eta$/$\gamma$/$\delta$, and $\epsilon$ ring groups. Each of the narrow main rings (6, 5, 4, $\alpha$, $\beta$, $\eta$, $\gamma$, $\delta$, $\epsilon$) was projected separately (i.e. one 2-D model per ring, ignoring all other rings) onto a very high resolution (0.01$''$/pixel) 2-D grid with proper semimajor axis, eccentricity, inclination, and observing geometry\footnote{Ephemeris data were taken from JPL Horizons and the Planetary Ring Node.} using the \texttt{PyAstronomy} package in \texttt{Python}\footnote{https://github.com/sczesla/PyAstronomy}. These super-resolution model rings were convolved with the Gaussian synthesized beam of ALMA, then summed into a radial profile in the same way as the data (see previous paragraph). The result of this exercise was a 1-D pseudo-Gaussian profile of each ring on each observing date that accounted for its eccentricity. These ring profiles were weighted according to their equivalent widths at visible wavelengths as measured by stellar occultations \citep[][]{french86}, added together, and then scaled by eye to match the data to produce an a-priori model radial profile of the ring system. A suite of models was then produced in a Markov Chain Monte Carlo (MCMC) framework implemented using the Python \texttt{emcee} package\footnote{http://dfm.io/emcee/current/} \citep[][]{dfm13}, wherein the fluxes $F_{654}$, $F_{\alpha \beta}$, $F_{\eta \gamma \delta}$, and $F_\epsilon$ of the four ring groups were allowed to vary freely while the relative contribution from each ring within a group was set according to its visible-wavelength equivalent width \citep[][]{french86}.  The flux of the 1-D model profile $F_m$ at all radii $r$ was thus fully specified by the fluxes of the four ring groups. Letting $\theta = [F_{654}, F_{\alpha \beta}, F_{\eta \gamma \delta}, F_{\epsilon}]$, the likelihood function $\ln p$ was given by
\begin{equation}
	\label{lnlike}
	\ln p(F|r, \sigma, \theta) = -\frac{1}{2} \Sigma_n \Big[ (F_n - F_m(\theta))^2 \sigma_n^{-2} + \ln(\sigma_n^{-2}) \Big]
\end{equation}
where $\sigma_n^2$ was the variance of the measured fluxes at each radius $F_n$. The MCMC simulation produced posterior probability distributions of $F_{654}$, $F_{\alpha \beta}$, $F_{\eta \gamma \delta}$, and $F_\epsilon$, which are given in Appendix \ref{appendix_extended} Figures \ref{corner3}, \ref{corner4}, and \ref{corner6} for images at 3.1 mm, 2.1 mm, and 1.3 mm wavelengths, respectively. The probability distributions show that mild degeneracy between the fluxes of the ring groups is present, and was most apparent at 1.3 mm where the spatial resolution was poorest. The mean flux values of each group are given in Table \ref{table_fluxes}. The uncertainties given in that table are a quadrature sum of the standard deviation of the MCMC-derived probability distribution and the flux calibration error (Section \ref{section_data}). Those mean values were used to generate the best-fitting radial profiles shown in orange in Figure \ref{radialint}, and random sets of $\theta$ values drawn from the MCMC samples were used to produce the light gray model profiles shown in the same figure.

\begin{table}
	\begin{tabular}{|c|c|c||c|c|c|c||}
	Wavelength &  & $B_{obs}$ & \multicolumn{4}{c||}{Total Flux, $F_\nu$ (mJy)} \\ 
	(mm) & UT Date & (degrees) & 6/5/4 & $\alpha$/$\beta$ & $\eta$/$\gamma$/$\delta$ & $\epsilon$ \\
	\hline
	3.1 & 2017-12-03 & 36.8 & $1.29 \pm 0.15$ & $1.91 \pm 0.19$ & $2.77 \pm 0.22$ & $13.73 \pm 0.73$ \\
    2.1  & 2017-12-27 & 36.4 & $5.19 \pm 0.67$ & $6.49 \pm 0.80$ & $8.15 \pm 0.89$ & $36.0 \pm 2.2$ \\
	1.3  & 2018-09-13 & 43.7 & $35.5 \pm 2.4$ & $15.5 \pm 2.2$ & $44.4 \pm 3.2$ & $112.2 \pm 6.3$ \\
	\end{tabular}
	\caption{Total millimeter flux measurements of each ring group. $B_{obs}$ denotes ring opening angle with respect to the observer. Note that two observations were taken at 3.1 mm, on 2017-12-03 and 2017-12-06, and co-added.\label{table_fluxes}}
\end{table}

The lower resolution and SNR of the VLT Q2 image precluded retrieval of the $\epsilon$ ring flux from a radial profile. We instead measured the total flux within an elliptical annulus of 3 beam-widths thickness ($\sim$33 pixels) centered on the $\epsilon$ ring.  To minimize contamination from the bright planet, we masked the planet such that the pixels within two beam widths of the planet's outer edge were omitted from the calculated ring flux. The choice of the mask size was informed by line scans through the images along the ansae, where the contributions from the rings and planet were most cleanly separated.  Due to the viewing geometry, the PSFs from the ring system and the planet partly overlapped away from the ansae; the masking consequently removed more pixels from these portions of the contributing ellipse and potentially underestimated the flux of the rings. To evaluate the accuracy of our approach, we applied the same method to synthetic images with azimuthally-uniform rings of known brightness, blurred by the observed PSF of our images and corrupted by different manifestations of synthetic noise. These tests showed that neglecting any attempt at correcting for the planetary contribution overestimated the ring flux by nearly 20\%, while the simple mask tended to underestimate the ring flux by less than 10\%. We therefore conservatively estimate our measured ring system total flux is accurate to within 20\%.  Finally, we assumed the contribution of the $\epsilon$ ring to the total flux of the ring system was the same at 18.7 $\mu$m as at 3 mm to get an approximate $\epsilon$ ring flux.

We find the best fit to the 3.1 mm radial profile with total fluxes relative to the $\epsilon$ ring of $9.4 \pm 1.4$\%, $13.9 \pm 1.4$\%, and $20.2 \pm 1.6$\% for the 6/5/4, $\alpha$/$\beta$, and $\eta$/$\gamma$/$\delta$ groups. These values are close to the relative equivalent widths at visible wavelengths, which \citet{french86} give as 7\%, 16\%, and 15\%, respectively.  However, at shorter wavelengths we find the 6/5/4 and $\eta$/$\gamma$/$\delta$ groups to be much brighter than expected based on the equivalent width measurements, by factors of 2.1 and 1.5 at 2.1 mm and 4.4 and 2.7 at 1.3 mm. This discrepancy may be due to unmodeled contributions from diffuse millimeter-sized dust between the inner rings, a larger fraction of millimeter-sized grains in those ring groups than in the $\epsilon$ ring, or systematic errors due to imperfect subtraction of the PSF around the bright planet. The latter effect may be especially consequential at 1.3 mm due to the poorer resolution of our data at that frequency.

The ring is bathed by the cosmic microwave background (CMB) from every angle, except from the angles where it is blocked by Uranus.  The final reported brightness temperature of Uranus was corrected to account for the blocked CMB photons according to the prescription in \citet{depater14}, which gives correction factors of 1.02 K, 0.59 K, and 0.12 K at 3.1 mm, 2.1 mm, and 1.3 mm, respectively. If that CMB correction is undone (that is, Uranus is assumed to be fainter than it really is), then the CMB appears to reach the rings fully isotropically and thus be scattered isotropically. The interferometer is insensitive to scattered CMB photons from the rings since it, too, is bathed isotropically by the CMB; thus, the measured flux from the rings does not need to be corrected for the CMB \citep[][]{dunn05, zhang19}.

\section{Thermal Modeling of the $\epsilon$ Ring}
\label{section_model}

We converted the $\epsilon$ ring total flux measurements derived in Section \ref{section_radialint} into spectral radiance (hereafter, ``brightness'') units to produce a coarse spectrum of the ring. To do so, we simply divided the total flux measurements by the projected geometric area of the $\epsilon$ ring, which was computed as follows. Let the measured width of the ring at periapsis and apoapsis be denoted $w_p$ and $w_a$ \citep[19.7 km and 96.4 km for the $\epsilon$ ring][]{karkoschka01b}. The inner and outer edge of the ring are defined by two ellipses sharing one focus (the center of mass of Uranus) but with semimajor axes $a$ and eccentricities $e$ such that $w_p$ and $w_a$ take their measured values. The inner ellipse has $a_{inner} = a - (w_p + w_a)/4$ and the outer ellipse has $a_{outer} = a + (w_p + w_a)/4$. To keep the focus at the same location requires the equation $a_{outer} - c_{outer} = a_{inner} - c_{inner} + w_p$ to be satisfied, where $c = ae$ is the focal distance of the ellipse. The equation is satisfied when $c_{outer} - c_{inner} = (w_a - w_p)/2$. Thus the area is given by
\begin{equation}
A_{r} = \pi a_{outer}^2 \sqrt{1 - e_{outer}^2} - \pi a_{inner}^2 \sqrt{1 - e_{inner}^2}
\end{equation}
Projecting into the observing geometry and scaling to the observer-Uranus distance yields an angular area of $\Omega_{r,obs} = A_{r} \sin B / d^2 = 1.57\times10^{-12}$ steradians.

We compare these four brightness measurements to a simple thermal model modified from the NEATM model for near-earth asteroids \citep[][]{harris98} to constrain the properties of the $\epsilon$ ring. The model gives the measured temperature $T$ of a ring particle as
\begin{equation}
	\label{equation_neatm}
	T^4 = \frac{L_\odot (1 - A_B)}{4 \pi \bar{\eta} \sigma \epsilon d^2}
\end{equation}
where $L_\odot$ is the solar luminosity, $d$ is the heliocentric distance to the asteroid, $\epsilon$ is the bolometric emissivity of the asteroid, $A_B$ is the Bond albedo, $\sigma$ is the Boltzmann constant, and $\bar{\eta}$ is the ``beaming factor'', a catch-all correction factor for low-phase-angle beaming, rotation, thermal inertia, and geometry (see Section \ref{section_eta} for a discussion of how $\bar{\eta}$ differs from the standard NEATM model's $\eta$).
To translate this temperature into a brightness $S_\nu$ we use the equation
\begin{equation}
	\label{equation_snu}
	S_\nu = f(B) B(\nu, T)
\end{equation}
where $B(\nu, T)$ is the Planck function at frequency $\nu$ and temperature $T$, and $f(B)$ is the ring's ``fractional visible area'', the fraction of paths from the observer through the projected geometric ring area that strike a ring particle (as opposed to passing through to background space).

\begin{figure}
	\includegraphics[width = 1.0\textwidth]{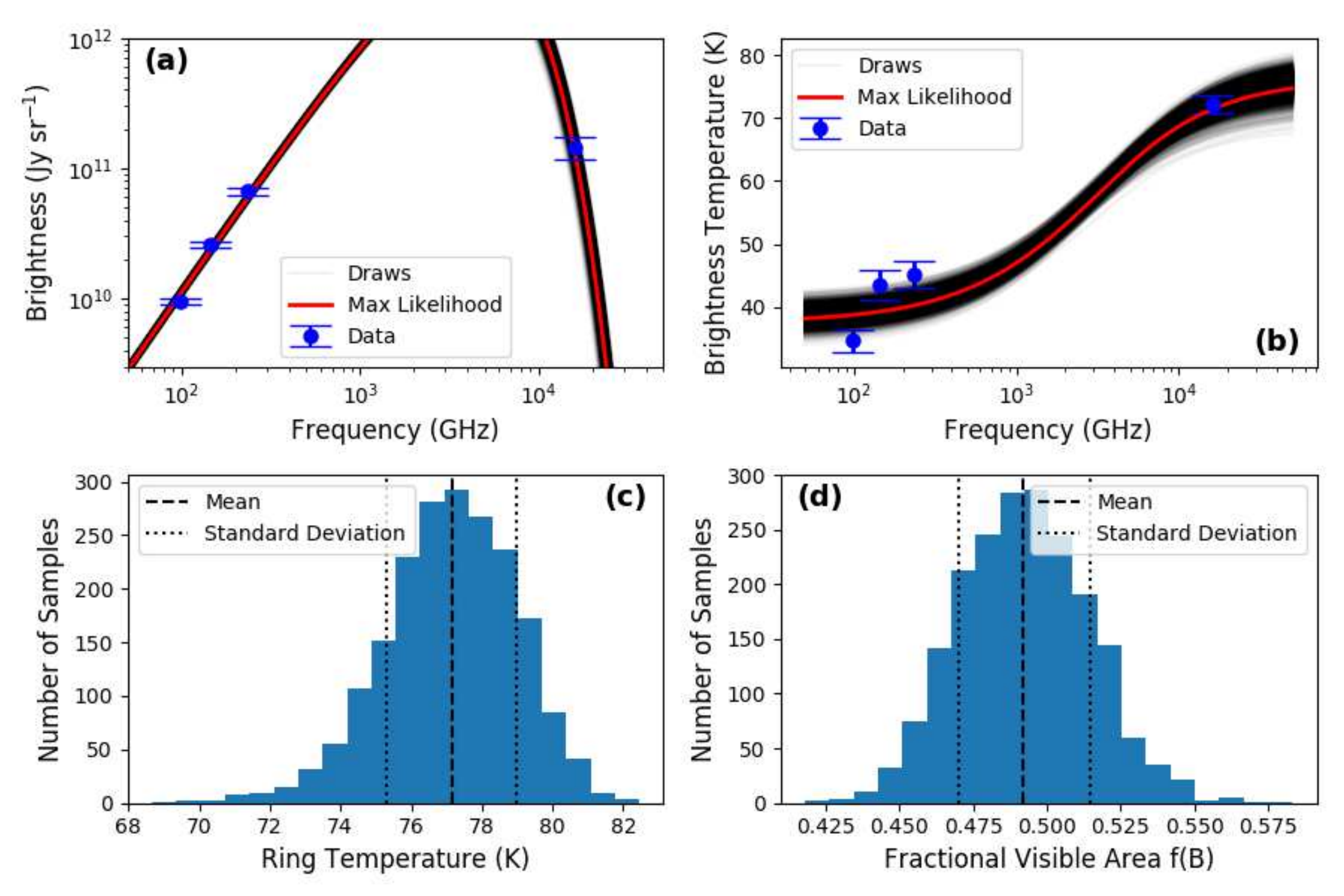}
	\caption{\textbf{(a)} Brightness and \textbf{(b)} brightness temperature comparison between retrieved NEATM model (red line: maximum-likelihood model, translucent black lines: random draws from probability distribution) and $\epsilon$ ring brightness measurements (blue points). \textbf{c)} Probability distribution function of $\epsilon$ ring temperature. \textbf{(d)} Probability distribution function of fractional visible area of $\epsilon$ ring. The mean value and quartiles are shown as dashed and dotted black lines, respectively.\label{neatm}}
\end{figure}

We performed a Markov Chain Monte Carlo (MCMC) sampling of our model's parameter space using the \texttt{emcee} Python package \citep[][]{dfm13} to explore the probability distributions of the free parameters in the thermal model. We allowed four parameters to vary: $\eta$, $\epsilon$, and $A_B$ in Equation \ref{equation_neatm}, and $f(B)$ in Equation \ref{equation_snu}. We treated $\eta$, $\epsilon$, and $f(B)$ as completely unknown, but provided a Gaussian prior on $A_B$ with mean value $0.061$ and standard deviation $0.006$ \citep[][]{karkoschka97}. In each MCMC iteration, a set of these parameters was drawn and plugged into the NEATM model, yielding a Planck function. The value of that Planck function at the ALMA- and VLT-observed frequencies was compared with the measured ring group brightness using a likelihood function similar in form to Equation \ref{lnlike}.  The simulation constrains the fractional visible area to $f(B) = 0.491 \pm 0.022$ and the particle temperature to $T = 77.3 \pm 1.8$ K; the best-fit model, as well as a suite of random model draws from the posterior distribution, are plotted against the observed $\epsilon$ ring total flux in Figure \ref{neatm}. That figure also shows posterior probability distributions for $f(B)$ and $T$, and a full ``corner plot'' displaying the one- and two-dimensional projections of the posterior probability distribution is shown in Appendix Figure \ref{corner}. The two figures show that $\bar{\eta}$ and the emissivity are completely degenerate with one another, and the retrieved albedo values are constrained only by the \citet{karkoschka97} prior; however, the fractional visible area $f(B)$ and temperature $T$ (derived by combining $\bar{\eta}$, $\epsilon$, and $A_B$) are strongly constrained and minimally degenerate with one another.

Our derived fractional visible area is consistent with the \citet{karkoschka01b} model, which gives $f(B) \approx 0.5$ at zero phase angle for a ring opening angle $B = 40^\circ$ ($B$ ranges from $37^\circ$ to $44^\circ$ for our observations). The emissivity of the ring cannot be constrained since it is degenerate with $\bar{\eta}$ and both are unknown a-priori. However, the product of the two quantities is determined to be $\bar{\eta} \epsilon = 1.62 \pm 0.17$. If we assume the ring particles are perfect graybody emitters such that $\epsilon = 1 - A_B$, then $\bar{\eta}$ takes a best-fit value of 1.72. This $\bar{\eta}$ value is larger, and the observed temperature is lower, than expected from the Standard Thermal Model for near-Earth asteroids \citep[STM;][]{lebofsky86}, which assumes very low thermal inertia and/or slow rotation such that all flux is emitted on the sun-facing side of the particle, leading to a predicted temperature $T_{STM} = 79.5$ K. However, $\bar{\eta}$ is smaller and $T$ higher than expected for a particle with thermal timescales long enough and rotation fast enough that the particle's temperature is independent of longitude as in the Fast-Rotating Model (FRM) \citep[][]{lebofsky89}, which yields $T_{FRM} = 71.5$ K (sub-solar latitude $\approx$40$^{\circ}$; see Section \ref{section_eta} for a description of our ring particle thermal model). This finding may indicate that the ring particles are closer to STM-like than FRM-like; that is, their thermal inertia is low enough, and their rotation rate is slow enough, that their dayside and nightside temperatures are different at a given latitude. However, beaming due to surface roughness tends to increase the amount of flux an observer sees, leading to lower $\bar{\eta}$ values and higher observed temperatures \citep[][and references within]{spencer89, lagerros98}. Surface roughness is not included in our model, so we cannot disentangle the effects of beaming from the effects of thermal inertia.

\subsection{Ring Particle Model}
\label{section_eta}

\begin{figure}
	\includegraphics[width = 0.7\textwidth]{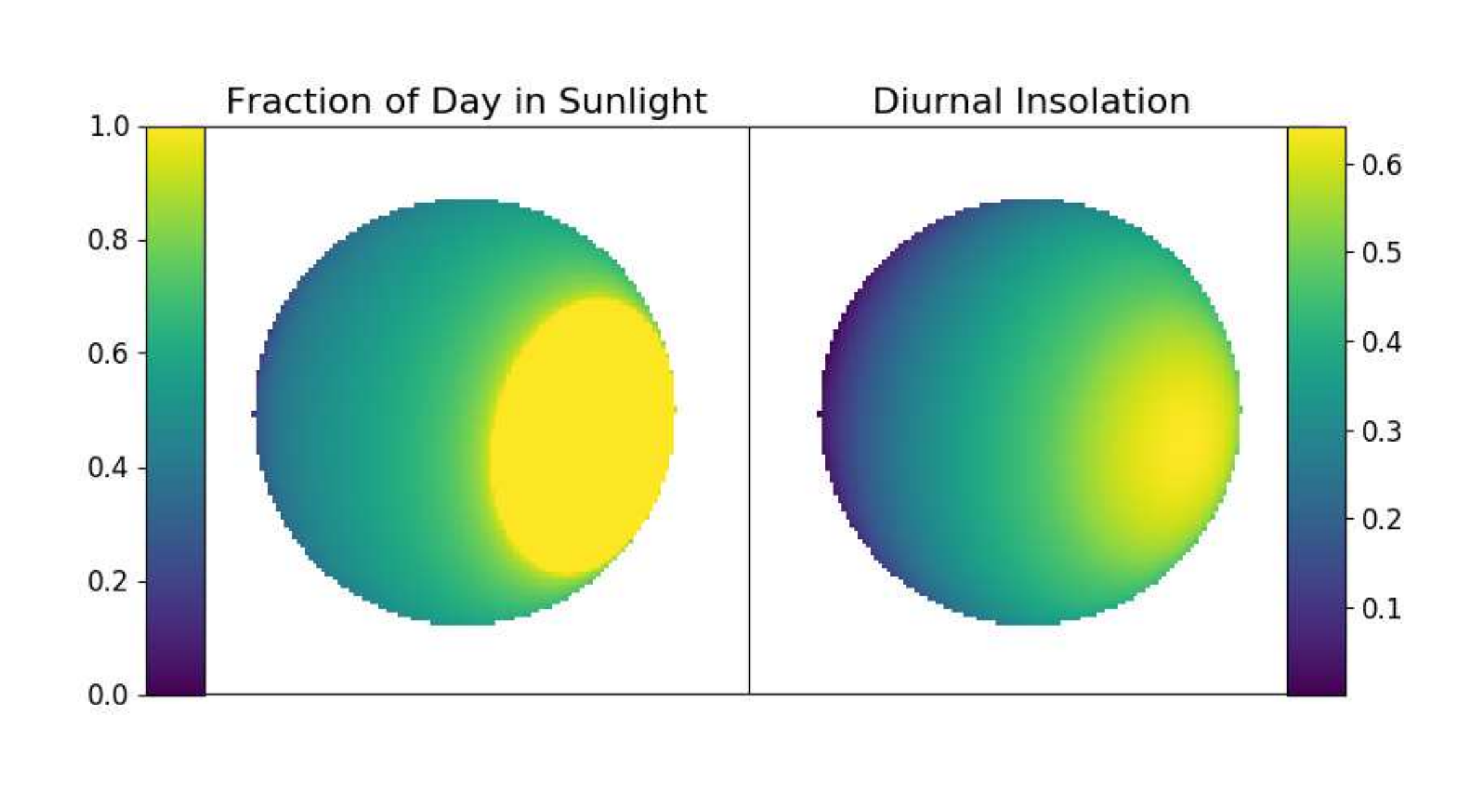}
	\caption{\textbf{Left:} Fraction $h_0/\pi$ of one rotation period that a model ring particle spends in sunlight. \textbf{Right:} Diurnal mean insolation $\bar{Q}$ of a model ring particle, normalized to the insolation received by a non-rotating (STM-like) model particle at the sub-solar point.\label{particlemodel}}
\end{figure}

The NEATM-like model we employ defines $\bar{\eta}$ as a property of the ring particles as a whole according to Equation \ref{equation_neatm}. This differs slightly from the original NEATM model \citep[][]{harris98}, which defines $\eta$ as a function of the temperature at the sub-solar point $T_0$ such that $T_0^4 \propto \eta$ and $T(\mu) = T_0 \mu^{1/4}$, where $\mu$ is the cosine of the emission angle. That is, $\bar{\eta}$ represents a geometric average while $\eta$ does not. To determine the physical meaning of our observed value of $\bar{\eta}$ for the $\epsilon$ ring, we produced a toy model of a ring particle. The model assumes a spherical particle with no surface roughness and no lateral heat conduction within the particle. The mean diurnal insolation $\bar{Q}$ for such a particle is given by
\begin{equation}
	\bar{Q} = \frac{L_\odot}{4 \pi d^2} \frac{1}{\pi} \Big( h_0 \sin \phi \sin \delta + \cos \phi \cos \delta \sin h_0 \Big)
\end{equation}
\citep[][]{pierrehumbert10}, where $\delta$ is the sub-solar latitude of the particle, $\phi$ is the latitude at a given location on the particle, and $h_0$ is the hour angle at sunrise and sunset in radians at a given location ($\cos h_0 = −\tan \phi \tan \delta$), which is equal to $\pi$ times the fraction of the day that location spends in sunlight (note that $h_0$ is set to $\pi$ at the summer pole and zero at the winter pole).  The temperature at each point is given by $T(\phi) = \bar{Q}^{1/4}$, and the total flux from the particle at the observed frequencies is found by applying a Planck function at each point on the model particle, then taking the mean.  We retrieve a blackbody temperature $T$ of the particle as a whole by fitting a single Planck function to these model flux values. Finally, $\bar{\eta}$ is found by comparing $T$ to the subsolar temperature of the STM at the same distance: $1 / \bar{\eta} = (T/T_{0,STM})^4$. Taking $T_{0, STM} = 88.3$ K as appropriate at Uranus's distance, this procedure yields $T = 79.5$ and $\bar{\eta} = 1.5$ for the STM as well as $T = 64.1$ and $\bar{\eta} = 3.6$ for the FRM. The FRM-derived temperature depends on the orientation of a particle's rotation axis with respect to the Sun, and the dense packing of the $\epsilon$ ring means its particles are likely to undergo frequent collisions that lead to a large spread in rotation axis orientations. However, the total angular momentum of the ring system should dictate that the average rotation axis is aligned with that of Uranus, so we assume this average rotation axis (sub-solar latitude $\approx$40$^\circ$) for all ring particles. Spatial variations in $h_0/\pi$ and $\bar{Q}$ are mapped in Figure \ref{particlemodel} for this geometry. In this case our model yields $T = 71.5$ and $\bar{\eta} = 2.3$. This temperature is lower, and $\bar{\eta}$ is higher, than retrieved from the data.

\section{Azimuthal Structure}
\label{section_az}

\begin{figure}
	\includegraphics[width = 0.7\textwidth]{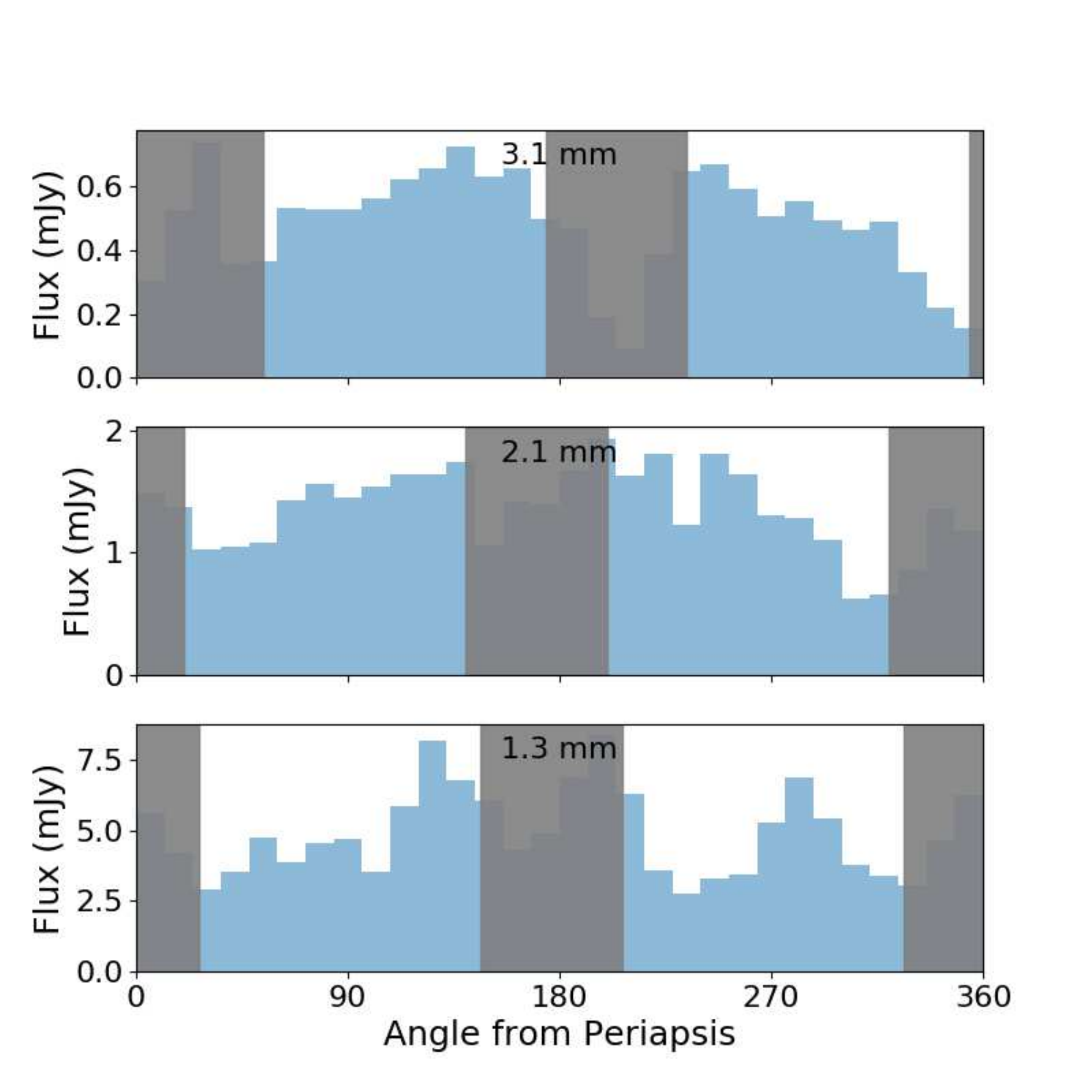}
	\caption{$\epsilon$ ring brightness as a function of azimuth. The grayed-out regions are contaminated by artifacts from the bright planet. A clear periapsis-apoapsis asymmetry is present at 3.1 mm and 2.1 mm wavelengths.\label{deproj}}
\end{figure}

We searched for azimuthal structure in the $\epsilon$ ring by breaking the ring into 30 bins, each subtending 12$^\circ$ of azimuth and having a width equal to two times the FWHM of the ALMA point-spread function. The measured brightness in these bins as a function of angle from periapsis is shown in Extended Data Figure \ref{deproj}.  We find the $\epsilon$ ring to be a factor of 2-3 brighter at apoapsis than at periapsis at 3.1 and 2.1 mm, in good agreement with stellar occultation measurements \citep[][]{french86}, visible/infrared reflected light observations \citep[][]{karkoschka97, depater02, depater13} and radio occultation measurements \citep[][]{gresh89}. No clear periapsis-apoapsis asymmetry is seen at 1.3 mm or 18.7 $\mu$m; however, the signal-to-noise ratio is lower in those data, and due to the poorer resolution of those data the measurements are contaminated by the presence of the inner rings to a greater extent. Note that we exclude the portions of the ring nearest (in projection) to Uranus due to contamination by imaging artifacts arising from the very millimeter-bright planet.


\section{Summary}
\label{section_summary}

Using ALMA and the VLT VISIR instrument, we have observed the thermal emission component of the Uranian ring system for the first time. Our observations of the $\epsilon$ ring agree very well with the \citet{karkoschka01b} model, supporting its derivation of the fractional visible area as a function of ring opening angle, and the 3.1 mm brightnesses of the inner main ring groups take the same ratio as observed at visible wavelengths. The consensus between our millimeter and mid-infrared observations and literature visible-wavelength observations shows that the properties of the main rings remain the same at any observed wavelength despite the fact that our observations are not sensitive to micron-sized dust. This finding confirms the hypothesis, proposed based on radio occultation results \citep[][]{gresh89}, that the main rings are composed of centimeter-sized or larger particles. A simple thermal model similar to the NEATM model for asteroids was applied to determine that the $\epsilon$ ring particles display roughly black-body behavior at millimeter/mid-infrared wavelengths at a temperature of $77 \pm 2$ K, suggesting the particles' thermal inertia may be small enough and rotation rate slow enough, to induce longitudinal temperature differences between their dayside and nightside. Observations at higher spatial resolution, achievable using ALMA, would resolve the inner main rings separately and quantify the millimeter-sized dust component if present.

\software{PyAstronomy, photutils, emcee, IDL} 

\acknowledgements

{\large\textit{Acknowledgements:}}

This paper makes use of the following ALMA data: ADS/JAO.ALMA\#2017.1.00855.S. ALMA is a partnership of ESO (representing its member states), NSF (USA) and NINS (Japan), together with NRC (Canada), MOST and ASIAA (Taiwan), and KASI (Republic of Korea), in cooperation with the Republic of Chile. The Joint ALMA Observatory is operated by ESO, AUI/NRAO and NAOJ. The National Radio Astronomy Observatory (NRAO) is a facility of the National Science Foundation operated under cooperative agreement by Associated Universities, Inc.

This investigation was partially based on thermal-infrared VISIR observations acquired at the ESO Very Large Telescope  Paranal UT3/Melipal Observatory, with program ID 0101.C-0073(B)

This research was supported in part by NASA Grant NNX16AK14G through the Solar System Observations (SSO) program
to the University of California, Berkeley.

E. Molter was supported in part by NRAO Student Observing Support grant \#SOSPA6-006.

Fletcher and Roman were supported by a European Research Council Consolidator Grant (under the European Union's Horizon 2020 research and innovation programme, grant agreement No 723890) at the University of Leicester.

This work made use of the PyAstronomy Python package.

We thank the staff at the 2018 NRAO Synthesis Imaging Workshop for providing advice on calibration of the ALMA data used in this paper.

\appendix

\setcounter{table}{0}
\renewcommand{\thetable}{A\arabic{table}}
\setcounter{figure}{0}
\renewcommand{\thefigure}{A\arabic{figure}}

\section{Scattered Light Contribution}
\label{section_rt}

Here we adopt a simple analytical radiative transfer calculation \citep[after][]{chandrasekhar60} to determine the contribution of scattered millimeter-wavelength light from Uranus to the millimeter brightness of the $\epsilon$ ring. The radiative transfer equation reads:
\begin{equation}
	\frac{dI_\nu}{d\tau_\nu} = S_\nu - I_\nu
\end{equation}
where $S_\nu = j_\nu /  \alpha_\nu$ is the source function. For scattered light from direction $(\theta', \phi')$ into direction $(\theta, \phi)$, the source function is given by
\begin{equation}
	S_\nu(\theta, \phi) = \varpi_\nu \int_{4\pi} \frac{p(\Theta)}{4\pi} I_\nu(\theta', \phi') d\Omega'
\end{equation}
where $p(\Theta)$ is the scattering phase function, $\Theta$ is the scattering angle, $\varpi_\nu$  is the single-scattering albedo, and $\int_{4\pi} I_\nu(\theta', \phi') d\Omega'$ is the flux density incident on the rings from the entire $4\pi$ steradian sky. Since light from Uranus is the only contributor to the incident radiation at the $\epsilon$ ring,
\begin{equation}
	\int_{4\pi} I_\nu(\theta', \phi') d\Omega' = I_{\nu,u} \Omega_{u,r}
\end{equation}
where $I_{\nu,u}$ is the specific intensity of Uranus at the observed frequency and $\Omega_{u,r}$ is the solid angle of Uranus as seen from the ring. Then the source function of the rings reads
\begin{equation}
	S_{\nu} = \varpi_\nu \frac{p(\Theta)}{4\pi} I_{\nu,u} \Omega_{u,r}
\end{equation}
To convert this into a flux density as seen from ALMA, we must multiply by the solid angle of the rings with respect to the observer $\Omega_{r,obs}$:
\begin{equation}
	\label{frobs}
	F_{\nu,r,obs} = \varpi_\nu \frac{p(\Theta)}{4\pi} I_{\nu,u} \Omega_{u,r} \Omega_{r,obs}
\end{equation}
The specific intensity is the same for any observer, so the same $I_{\nu,u}$ measured by ALMA would be measured by an observer at the $\epsilon$ ring, and thus
\begin{equation}
	\label{iu}
	I_{\nu,u} = \frac{F_{\nu,u,r}}{\Omega_{u,r}} = \frac{F_{\nu,u,obs}}{\Omega_{u,obs}}
\end{equation}
where $F_{\nu,u,r}$ and $F_{\nu,u,obs}$ denote the spectral flux density of Uranus as seen from the rings and from Earth, respectively. Substituting Equation \ref{iu} into Equation \ref{frobs}, we have
\begin{equation}
	F_{\nu,r,obs} = \varpi_\nu \frac{p(\Theta)}{4\pi} \frac{\Omega_{u,r} \Omega_{r,obs}}{\Omega_{u,obs}} F_{\nu,u,obs}
\end{equation}
The solid angle is defined as $\Omega = A/d^2$, where $A$ is the projected area of an object and $d$ is the distance from that object to the observer, so
\begin{equation}
	\frac{\Omega_{u,r}}{\Omega_{u,obs}} = \frac{A_u d_{obs}^2}{A_u r_r^2} = \frac{d_{obs}^2}{r_r^2}
\end{equation}
where $A_u$ is the cross-sectional area of Uranus and $r_r$ is the distance from Uranus to the $\epsilon$ ring. Combining this with the fact that $\Omega_{r,obs} = A_{r,obs}/d_{obs}^2$, where $A_{r,obs}$ is the cross sectional area of the $\epsilon$ ring with respect to the observer, we end up with
\begin{equation}
	\label{f}
	F_{\nu,r,obs} = \varpi_\nu \frac{p(\Theta)}{4\pi} \frac{A_{r,obs}}{r_r^2} F_{\nu,u,obs}
\end{equation}

This derivation assumed that the entire area of the ring is illuminated by light from Uranus as seen by the observer. However, due to shadowing and gaps in the ring, only a fraction of the ring $f(B_u, B)$ is illuminated, where $B_u$ denotes the ring opening angle for an observer on Uranus and $B$ denotes the ring opening angle for an observer on Earth. In reality the flux comes from a distribution of angles: the diameter of Uranus is roughly $52.6^\circ$ at the distance to the $\epsilon$ ring, giving ring opening angles between $-26.3^\circ$ and $26.3^\circ$.  To evaluate $f(B_u, B)$ precisely is a difficult task, since one must integrate over many Uranus-ring-observer angles. To simplify the problem we shall make two assumptions. First, we assume $B_u$ is well represented by the geometric average opening angle $B_u = 18.4^{\circ}$ (the orbital inclination of the ring with respect to Uranus is $i = 0.001^\circ$, which is negligible). Second, we assume that half of the ring area illuminated by light from Uranus can be seen from Earth, to account for surfaces that point away from the observer. Under these assumptions, $f(B_u, B) = f(B_u)/2$, and we can again use the \citet{karkoschka01b} prescription to calculate $f(B_u)$. Since the photons originate from Uranus itself the phase angle is $0^\circ$ and the model gives us $f(B_u, B) = f(18.4)/2 = 0.18$. The physical interpretation of such a low value is that at small incident angles particles cast long shadows on each other, blocking much of the flux. Applying this correction to Equation \ref{f} gives us
\begin{equation}
	\label{ffinal}
	F_{\nu,r,obs} = f(B_u, B) \varpi \frac{p(\Theta)}{4\pi} \frac{A_{r,obs}}{r_r^2} F_{\nu,u,obs}
\end{equation}

\noindent Since the particles in the $\epsilon$ ring are assumed to be $\gtrsim$10 cm in diameter \citep[][]{gresh89, karkoschka01b}, we assume a phase function $p(\Theta) = 1$ appropriate for geometric scattering.

Our ALMA observations give the total 3 mm flux density of Uranus to be $F_{\nu,u,obs} = 8.80$ Jy. We assume the ring particles have albedo $\varpi_{3mm} = 1 - \epsilon_{3mm} = 0.05$. For simplicity we take $r_r = 51148$ km, the average of the $\epsilon$ ring semimajor and semiminor axes. At ring opening angle $B_{obs} = 36.77^\circ$, the projected ring area is $A \sin B_{obs} = 1.11\times10^7$ km$^2$. Plugging these numbers into Equation \ref{ffinal} yields $F_{3mm,r,obs} = 28$ $\mu$Jy.
This value is a factor of $\sim$500 smaller than the measured ring flux at 3.1 mm. Since both the planet and rings are at least relatively near the same temperature, the ratio between the reflected thermal emission from the planet and the direct thermal emission from the rings is nearly wavelength-independent, and similarly small values are found in the mid-infrared.  While many simplifying assumptions were made in this derivation, such a small scattered light contribution means that scattered light from Uranus can be ignored.

Equation \ref{ffinal} can be applied to quantify the contribution of scattered radio flux from the Sun with the following modifications. The fraction $f(B_\odot, B)$ can once again be taken from the \citet{karkoschka01b} model, where $B_\odot$ is the ring opening angle as viewed from the Sun (in fact, this geometry is the scenario for which the model was developed). Instead of the Uranus-ring distance $r_r$, the distance from the Sun to the ring $d_{\odot,r}$ should be used. Finally, the Sun's radio flux at Earth $F_{\nu,\odot,obs}$ replaces $F_{\nu,u,obs}$. Thus we have
\begin{equation}
	F_{\nu,r,obs} = f(B_\odot, B) \varpi \frac{p(\Theta)}{4\pi} \frac{A_{r,obs}}{d_{\odot,r}^2} F_{\nu,\odot, obs}
\end{equation}
At the time of our 3.1 mm observations, the phase angle was very nearly 2$^\circ$, so the \citet{karkoschka01b} model gives $f(B_\odot, B) = 0.5$. \citet{alissandrakis17} used ALMA observations to find a solar brightness temperature of $7250 \pm 170$ K at 100 GHz, as well as a solar radius of $964.1 \pm 4.5''$ at 1 AU at 100 GHz. Assuming the same temperature and radius at 97.5 GHz (3.1 mm), we get $F_{\nu,\odot,obs} = 1.45\times10^8$ Jy. Finally, taking $d_{\odot,r}$ at the time of observation from JPL Horizons, we find $F_{\nu,r,obs} = 0.38$ $\mu$Jy, which is $\sim$0.002\% of the measured flux. At 18 $\mu$m wavelengths, we find $F_{\nu,r,obs} = 7.8$ mJy, or $\sim$2\% of the measured flux; this is much smaller than the 20\% VLT flux calibration error and so is ignored.




\clearpage
\section{Extended Data}
\label{appendix_extended}

\begin{figure}[H]
	\includegraphics[width = 1.0\textwidth]{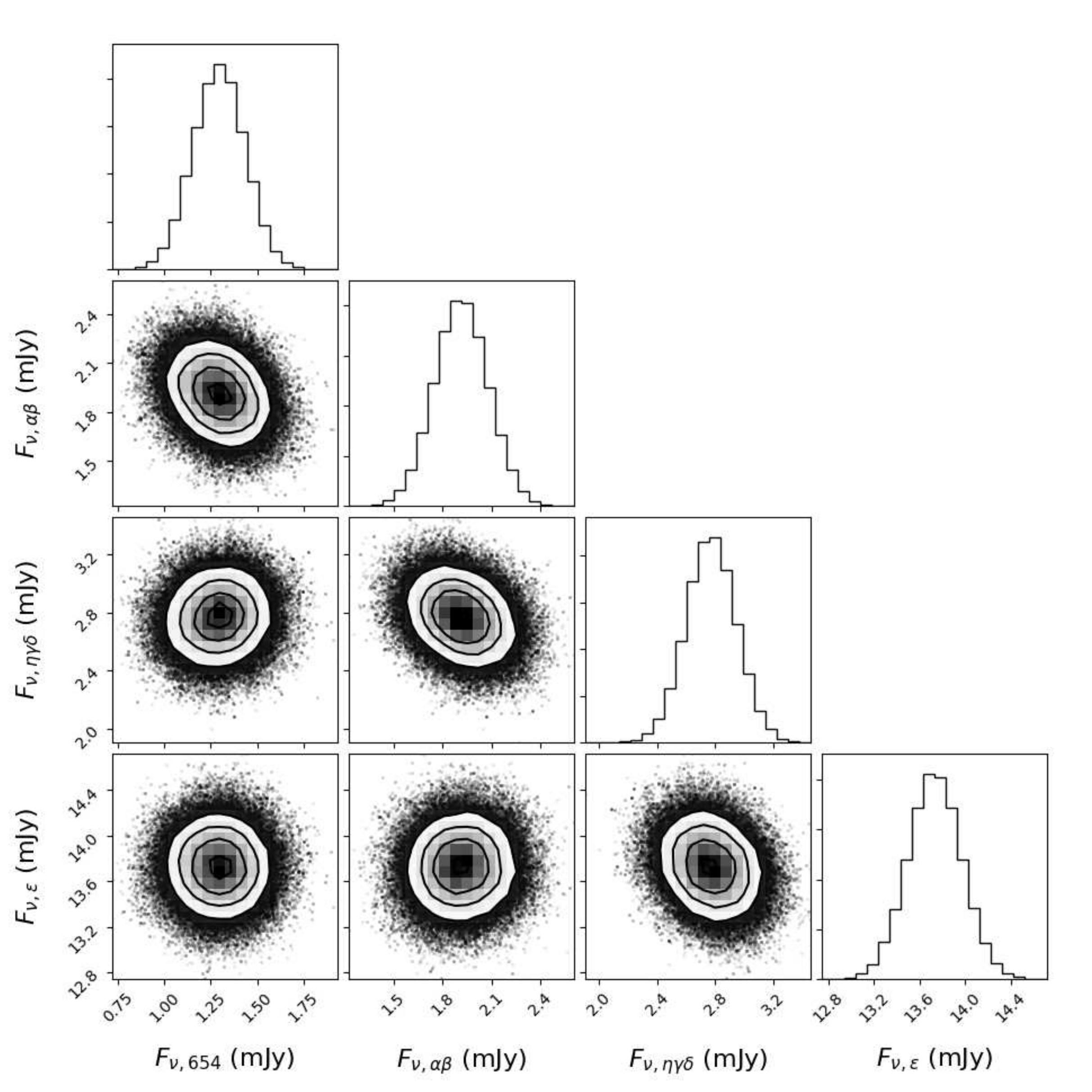}
	\caption{``Corner plot'' showing the one- and two-dimensional projections of the posterior probability distributions of the MCMC-retrieved total fluxes of each ring group in Band 3 (3.1 mm), derived from comparing our model to the observed the radial profile.\label{corner3}}
\end{figure}

\begin{figure}[H]
	\includegraphics[width = 1.0\textwidth]{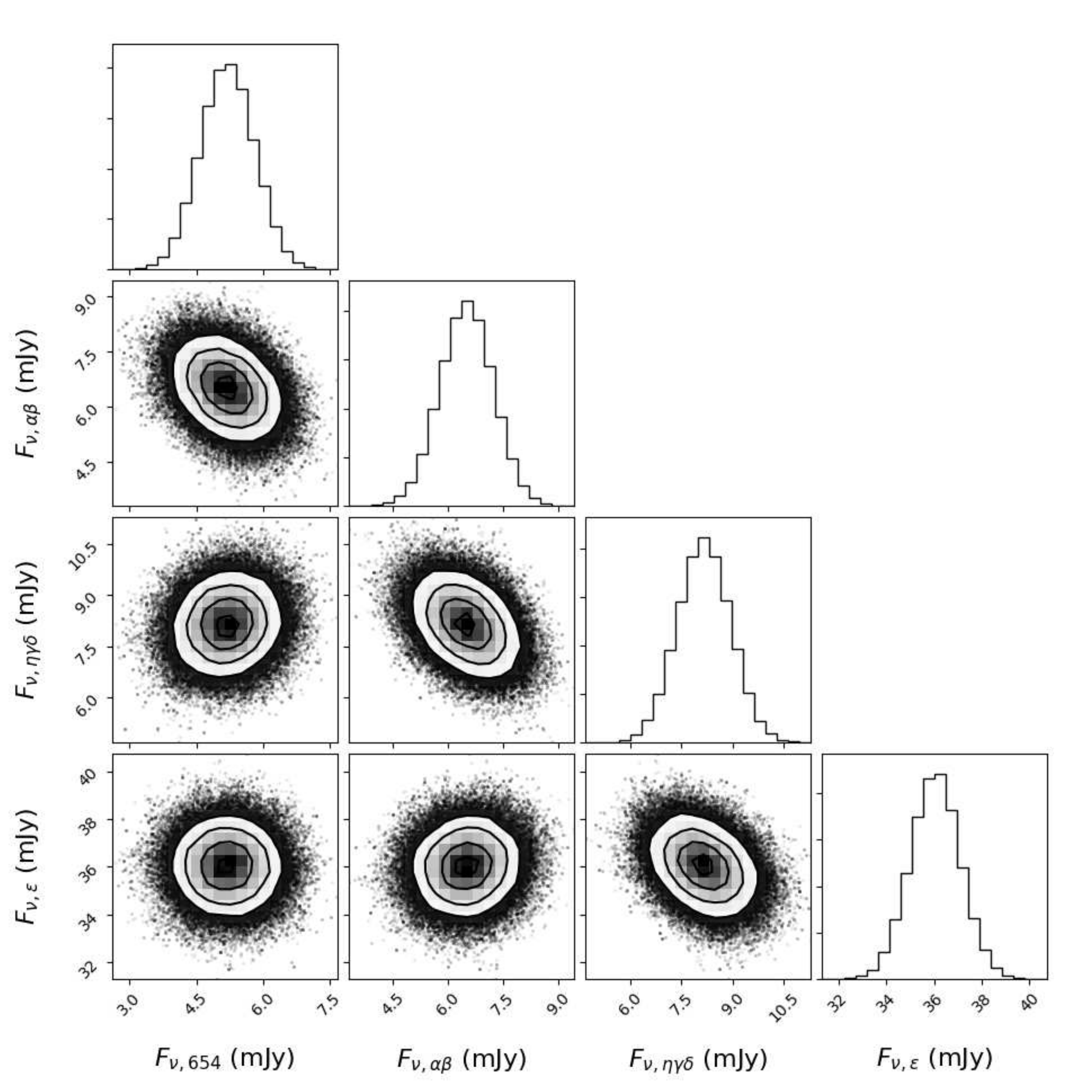}
	\caption{``Corner plot'' showing the one- and two-dimensional projections of the posterior probability distributions of the MCMC-retrieved total fluxes of each ring group in Band 4 (2.1 mm), derived from comparing our model to the observed the radial profile.\label{corner4}}
\end{figure}

\begin{figure}[H]
	\includegraphics[width = 1.0\textwidth]{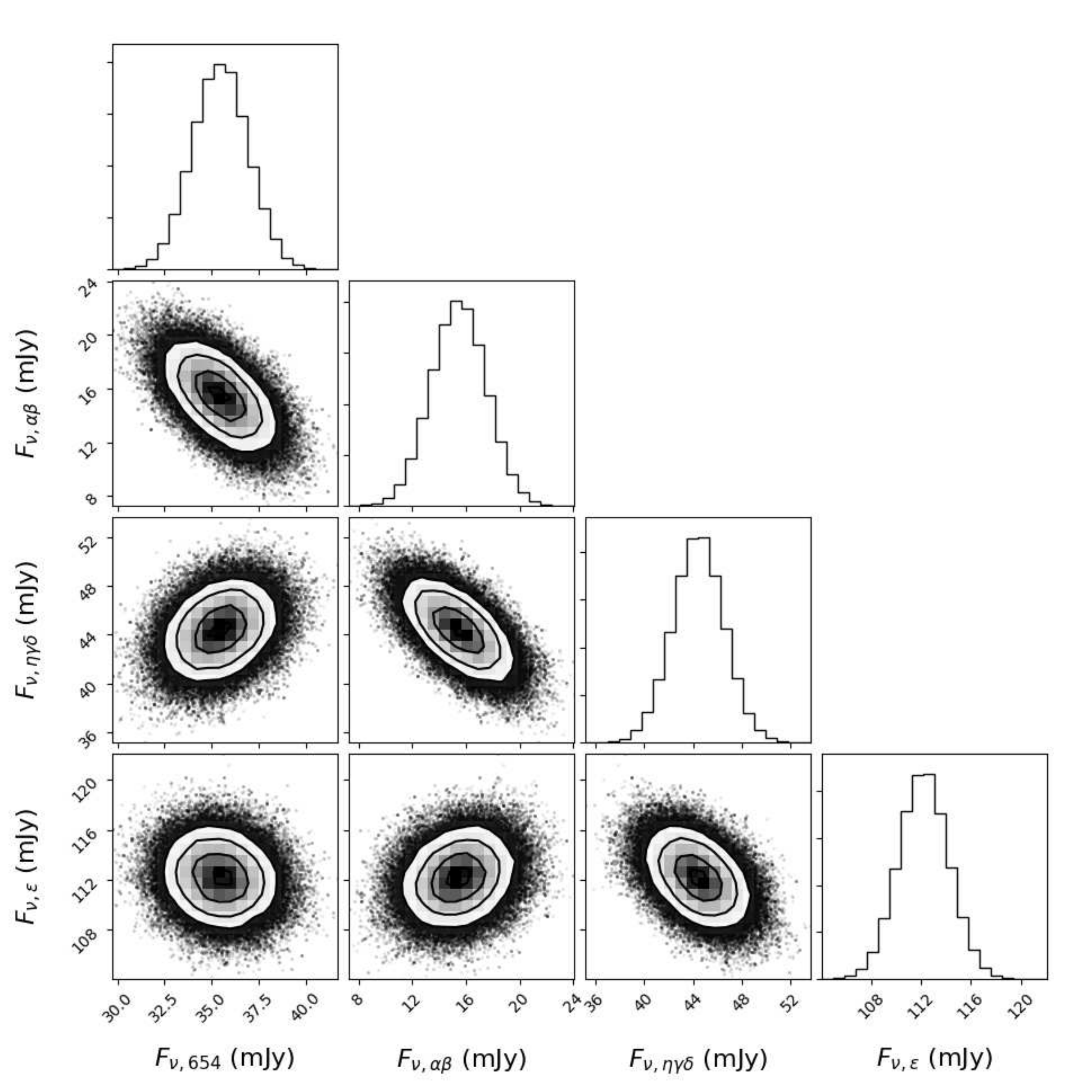}
	\caption{``Corner plot'' showing the one- and two-dimensional projections of the posterior probability distributions of the MCMC-retrieved total fluxes of each ring group in Band 6 (1.3 mm), derived from comparing our model to the observed the radial profile.\label{corner6}}
\end{figure}

\begin{figure}[H]
	\includegraphics[width = 1.0\textwidth]{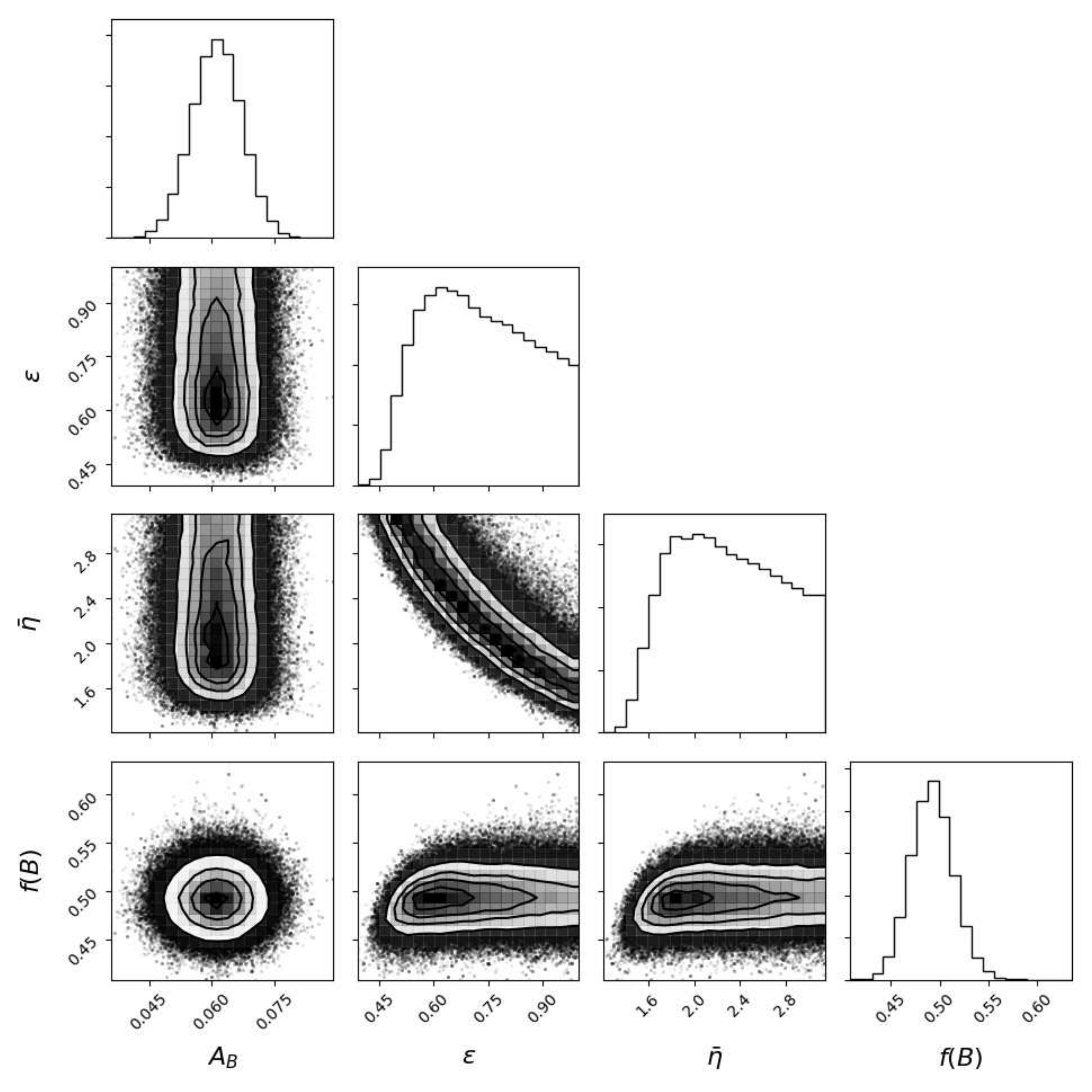}
	\caption{``Corner plot'' showing the one- and two-dimensional projections of the posterior probability distributions of the MCMC-retrieved thermal model parameters.\label{corner}}
\end{figure}

Figures \ref{corner3}, \ref{corner4}, and \ref{corner6} display ``corner plots'' for the MCMC-derived total flux measurements discussed in Section \ref{section_radialint}. Figure \ref{corner} displays a corner plot for the thermal model discussed in Section \ref{section_model}.

\clearpage
\bibliography{paper_aj}{}

\end{document}